\documentclass[12pt]{article}
\usepackage{setspace}
\usepackage{epsfig}
\usepackage{amssymb}
\def\be{\begin{equation}}
\def\ee{\end{equation}}
\def\ba{\begin{array}}
\def\ea{\end{array}}
\def\beqn{\begin{eqnarray}}
\def\eeqn{\end{eqnarray}}

\def\bt{\begin{tabular}}
\def\et{\end{tabular}}
\def\bc{\begin{center}}
\def\ec{\end{center}}
\newcommand{\mat}[9] {\left( \matrix{#1 & #2 & #3 \cr
                                     #4 & #5 & #6 \cr
                                     #7 & #8 & #9 \cr} \right)}
\begin{document}
\title{Constructing the Leptonic Unitarity Triangle}
\author{Gulsheen Ahuja and Manmohan Gupta\\
\\
{\it Department of Physics, Centre of Advanced Study, P.U.,
Chandigarh, India.}\\{\it Email: mmgupta@pu.ac.in}}
 \maketitle
\begin{abstract}
Following analogy of the `$db$' triangle in the quark mixing case,
we have explored the construction of the `$\nu_2.\nu_3$' leptonic
unitarity triangle using the MNS matrix obtained by Bjorken {\it
et al.} through generalization of the tribimaximal scenario. In
particular, for the $U_{e3}$ range $0.05-0.15$, the existence of
leptonic unitarity triangle indicates a fairly good possibility of
having non zero CP violation.
 \end{abstract}
In the last few years, apart from establishing the hypothesis of
neutrino oscillations, impressive advances have been made in
understanding the phenomenology of neutrino oscillations through
solar neutrino experiments \cite{solexp}, atmospheric neutrino
experiments \cite{atmexp}, reactor based experiments
\cite{reacexp} and accelerator based experiments \cite{accexp}
enabling the determination of the basic form of the
Maki-Nakagawa-Sakata (MNS) leptonic mixing matrix \cite{mns}. At
present, one of the key issues in the context of neutrino
oscillation phenomenology is the existence of CP violation in the
leptonic sector.

Taking clues from the existence of the unitarity triangle and
consequently CP violation in the quark sector \cite{quarkuni},
several attempts \cite{lepuni}-\cite{ bjorken} have been made to
explore such a possibility in the leptonic sector. In particular,
Farzan and Smirnov \cite{smirnov} have discussed the desirability
of exploring the construction of leptonic unitarity triangle for
finding possible clues to the existence of CP violation in the
leptonic sector. Considering the `$e$-$\mu$' triangle, for
$U_{e3}$ values in the range $0.09-0.22$, they have examined the
detailed implications of different values of CP violating phase
$\delta$ on the possible accuracy required in the measurement of
various oscillation probabilities. Very recently, Bjorken {\it et
al.} \cite{bjorken}, have constructed a generalization of the
tribimaximal scenario and have not only presented a very useful
form of the MNS matrix but have also proposed a unitarity
triangle, referred to as `$\nu_2.\nu_3$' which could be leptonic
analogue of the much talked about `$db$' triangle in the quark
sector. This immediately suggests a need for deeper study of the
`$\nu_2.\nu_3$' triangle using clues from the `$db$' triangle in
the quark sector. In particular, it would be very much desirable,
as a complimentary approach to the scenario investigated by Farzan
and Smirnov \cite{smirnov}, to find the probable values of CP
violating phase $\delta$ suggested by generalized tribimaximal
scenario of Bjorken {\it et al.} \cite{bjorken}.

To this end, taking clues from `$db$' unitarity triangle in the
quark sector, the purpose of the present paper is to explore the
possibility of the construction of the leptonic unitarity triangle
as well as the existence of CP violation in the leptonic sector.
In particular, in the MNS matrix constructed by Bjorken {\it et
al.} \cite{bjorken} we have considered different values of
$U_{e3}$ suggested by various theoretical models \cite{albright}.
Further, we have explored in detail the possibility of finding a
non zero value of $J_l$, the Jarlskog's rephasing invariant
parameter in the leptonic sector as well as the related Dirac-like
CP violating phase $\delta$.

For ready reference as well as to facilitate discussion of
results, we begin with the neutrino mixing phenomenon expressed in
terms of a $3 \times 3$ neutrino mixing MNS matrix \cite{mns}
given by \be \left( \ba{c} \nu_e \\ \nu_{\mu} \\ \nu_{\tau} \ea
\right)
  = \left( \ba{ccc} U_{e1} & U_{e2} & U_{e3} \\ U_{\mu 1} & U_{\mu 2} &
  U_{\mu 3} \\ U_{\tau 1} & U_{\tau 2} & U_{\tau 3} \ea \right)
 \left( \ba {c} \nu_1\\ \nu_2 \\ \nu_3 \ea \right),  \label{nm}  \ee
where $ \nu_{e}$, $ \nu_{\mu}$, $ \nu_{\tau}$ are the flavor
eigenstates and $ \nu_1$, $ \nu_2$, $ \nu_3$ are the mass
eigenstates. Following PDG representation, involving three angles
and the Dirac-like CP violating phase $\delta$ as well as the two
Majorana phases $\alpha_1$, $\alpha_2$, the MNS matrix $U$ can be
written as \beqn U ={\left( \ba{ccl} c_{12} c_{13} & s_{12} c_{13}
& s_{13}e^{-i \delta} \\ - s_{12} c_{23} - c_{12} s_{23} s_{13}
e^{i \delta} & c_{12} c_{23} - s_{12} s_{23} s_{13} e^{i \delta} &
s_{23} c_{13}
\\ s_{12} s_{23} - c_{12} c_{23} s_{13} e^{i \delta} & - c_{12}
s_{23} - s_{12} c_{23} s_{13} e^{i \delta} & c_{23} c_{13} \ea
\right)} \left( \ba{ccc} e^{i \alpha_1/2} & 0 & 0 \\ 0 &e^{i
\alpha_2/2} & 0 \\ 0 & 0  & 1 \ea \right). \label{nmm} \eeqn The
Majorana phases $\alpha_1$ and $\alpha_2$ do not play any role in
neutrino oscillations and henceforth would be dropped from the
discussion.

Unitarity implies nine relations, three in terms of normalization
conditions, the other six can be defined as
 \be \sum_{i=1,2,3} U_{\alpha i} {U^*_{\beta i}} =\delta_{\alpha\beta}
 ~~~~~~~(\alpha \neq \beta) , \label{ut1} \ee
 \be \sum_{\alpha=e,\mu,\tau} U_{\alpha i}
{U^*_{\alpha j}} =\delta_{ij}  ~~~~~~~(i \neq j) , \label{ut2} \ee
where Latin indices run over the mass eigenstates $(1,2,3)$ and
Greek ones run over the flavor eigenstates $(e,\mu,\tau)$. These
six non-diagonal relations can also be expressed through six
independent unitarity triangles in the complex plane.

\begin{figure}[tbp]
\centerline{\epsfysize=2.4in\epsffile{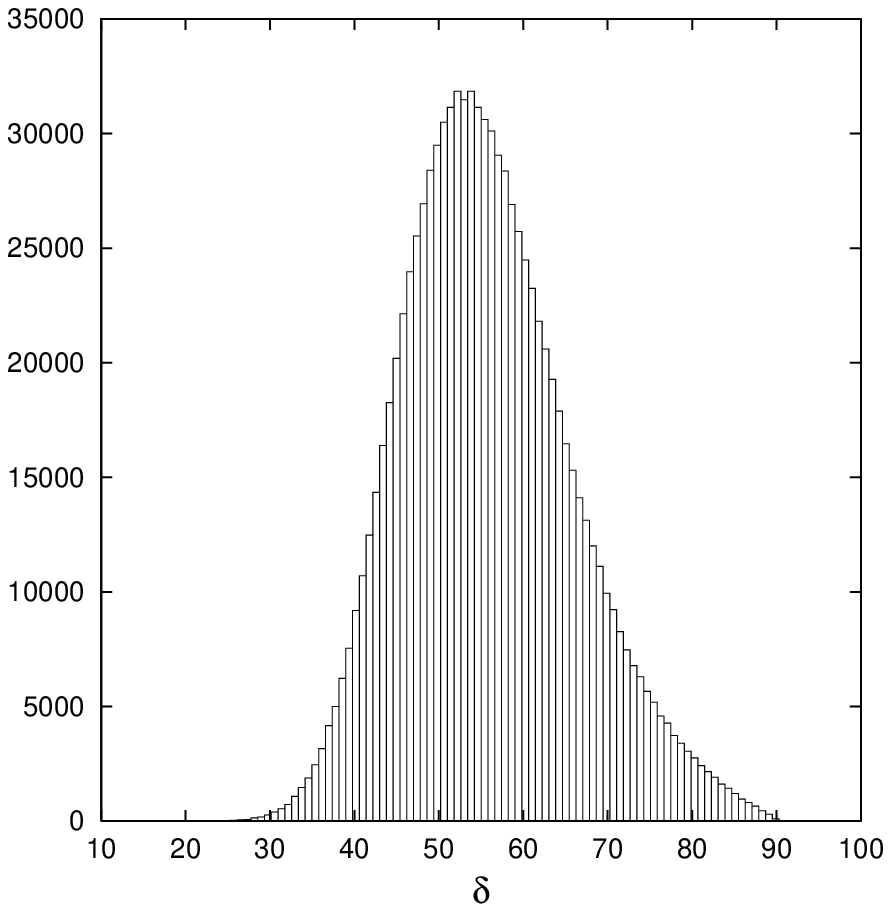}} \vspace{0.08in}
   \caption{Histogram of $\delta$ plotted by considering `$db$'
triangle in the case of quarks}
  \label{delqua}
  \end{figure}

For getting viable clues to the construction of the leptonic
unitarity triangle, we first consider the case of quarks wherein
the CKM matrix \cite{ckm} is fairly well established as well as
the CP violating phase $\delta$ has also been measured recently
\cite{pdgnew}-\cite{hfag} with a good deal of accuracy. To begin
with, we consider the quark mixing matrix given by PDG 2006
\cite{pdgnew} and attempt to reconstruct the CP violating phase
$\delta$ using the Jarlskog's rephasing invariant parameter $J$,
equal to twice the area of any of the unitarity triangle. In this
context, we consider the usual `$db$' triangle, expressed as
  \be V_{ud} {V_{ub}}^* + V_{cd} {V_{cb}}^* + V_{td}
{V_{tb}}^* =0\,,\label{db} \ee from which one can obtain a
histogram of $J$ by considering Gaussian distribution for the CKM
matrix elements as well as  imposing the constraints $|a|+|c|>
|b|$ and $|b|+|c|> |a|$ for the three sides of the triangle $a$,
$b$, $c$. From the histogram of $J$, not shown here, one can find
\be
J= (3.0 \pm 0.4) \times 10^{-5}.\ee Using the relation between the
parameter $J$ and phase $\delta$, e.g.,
\be
J=s_{12}s_{23}s_{13}c_{12}c_{23}c_{13}^2 \,{\rm sin}\,\delta,
\label{jd}\ee one can obtain the corresponding histogram of
$\delta$, shown in figure \ref{delqua}, yielding \be
\delta=55.4^{\rm o} \pm 10.0^{\rm o}. \label{delq} \ee For further
details we refer the reader to \cite{monpro}. Interestingly, we
find that the above mentioned $J$ value has an excellent overlap
with that found by PDG through their recent global analysis
\cite{pdgnew}. Also, this value of $\delta$ is fully compatible
with the experimentally determined $\delta$ given by PDG 2006 as
well as found by some of the most recent analyses
\cite{pdgnew}-\cite{hfag}.

The above discussion immediately provides a clue for exploring the
possibility of existence of CP violation in the leptonic sector,
even when leptonic mixing matrix is approximately known. In this
context, we have considered the MNS matrix obtained by Bjorken
{\it et al.} \cite{bjorken}, e.g.,
\be
U= X ~Y \label{uxy} \ee  where
\begin{eqnarray}
 X= \left( \matrix{
  \frac{2}{\sqrt{6}} &  \frac{1}{\sqrt{3}}  &   0  \cr
 -\frac{1}{\sqrt{6}} &  \frac{1}{\sqrt{3}}  & \frac{1}{\sqrt{2}}   \cr
 -\frac{1}{\sqrt{6}} &  \frac{1}{\sqrt{3}}  & -\frac{1}{\sqrt{2}}    \cr
 }\right),~~~~~~~~Y=\mat
  {C}{0}{\sqrt{\frac{3}{2}}U_{e3}}
  {0}{1}{0}
  {-\sqrt{\frac{3}{2}}U^*_{e3}}{0}{C}\end{eqnarray}
with
\begin{equation}
C=\sqrt{1-\frac{3}{2}|U_{e3}|^2}. \label{cdef}
\end{equation}
The matrix $X$ corresponds to the tribimaximal mixing matrix
whereas the matrix $Y$ breaks this exact tribimaximal form by
small perturbations due to the effects of the element $U_{e3}$. It
may be added that the matrix $U$ is unitary by construction.

It needs to be mentioned that the experimental data deviates
somewhat from the tribimaximal form, therefore we have modified
the matrix $X$ by introducing modest error of 5$\%$ to its `$12$'
and `$23$' elements, governed by solar and atmospheric neutrino
data. These elements have been considered independent in the
present case, the other elements along with the respective errors
have been obtained by using the constraints of unitarity. For
$U_{e3}$, it is very well recognized that its value would have
deep implications for the neutrino oscillation phenomenology
\cite{mnsm}-\cite{minakata}. However at present only its upper
limit is known, therefore while constructing the matrix $Y$ we
have taken its values to be 0.05, 0.10 and 0.15 and attached
10$\%$ errors. These values have been considered primarily
following a recent detailed analysis by Albright and Chen
\cite{albright} wherein they have studied the implications of
$U_{e3}$ values on various leptonic and grand unified models of
neutrino masses and mixings. The errors in the elements of the
matrices $X$ and $Y$ have been introduced following Farzan and
Smirnov \cite{smirnov} and keeping in mind the accuracy with which
these would be measured by the planned neutrino experiments
\cite{smirnov}, \cite{white}, \cite{marciano}- \cite{sugiyama}.
The matrices corresponding to $U_{e3}$ values $0.05 \pm 0.005$,
$0.10 \pm 0.01$ and $0.15 \pm 0.015$ are respectively as follows
 \be U = \left(  \ba{ccc}
  0.8150\pm0.0204 & 0.5774\pm0.0289  &  0.05\pm0.005\\
0.4508\pm0.0646  &  0.5774\pm0.0144  & 0.6808\pm0.0356\\
 0.3642\pm0.0646 &0.5774\pm0.0144 &0.7308\pm0.0356
 \ea \right), \label{.05} \ee
 \be U = \left(  \ba{ccc}
  0.8104\pm0.0203 & 0.5774\pm0.0289  &  0.1\pm0.01\\
0.4918\pm0.0648  &  0.5774\pm0.0144  & 0.6518\pm0.0363\\
 0.3186\pm0.0648 &0.5774\pm0.0144 &0.7518\pm0.0363
 \ea \right), \label{.1} \ee
 \be U = \left(  \ba{ccc}
  0.8026\pm0.0202 & 0.5774\pm0.0289  &  0.15\pm0.015\\
0.5312\pm0.0652  &  0.5774\pm0.0144  & 0.6201\pm0.0376\\
 0.2714\pm0.0652 &0.5774\pm0.0144 &0.7701\pm0.0376
 \ea \right), \label{.15} \ee
wherein we have given the magnitude of the elements, as is usual.

\begin{figure}[tbp]
\centerline{\epsfysize=2.4in\epsffile{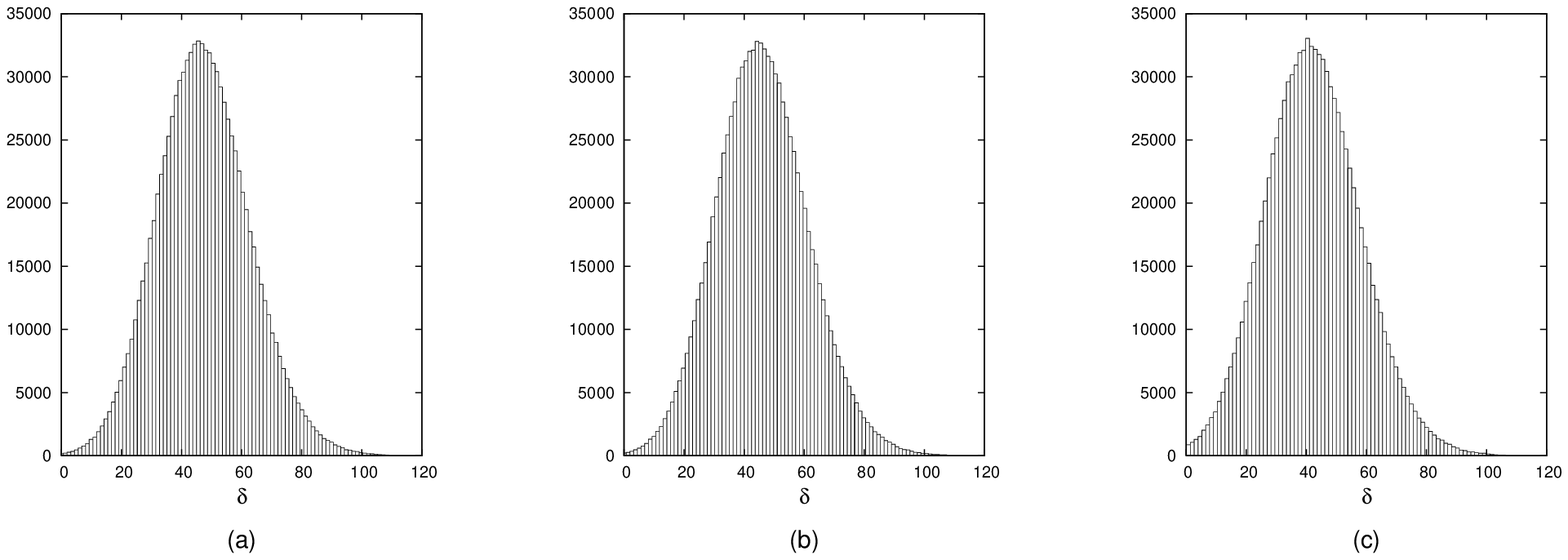}}
\vspace{0.08in} \caption{Histogram of $\delta$ plotted by
considering `$\nu_2.\nu_3$' triangle in the case of neutrinos for
(a) $U_{e3}=0.05 \pm 0.005$ (b) $U_{e3}=0.1 \pm 0.01$ (c)
$U_{e3}=0.15 \pm 0.015$ }
  \label{delneut}
  \end{figure}

Out of the six triangles defined by equations (\ref{ut1}) and
(\ref{ut2}), Bjorken {\it et al.} \cite{bjorken} have considered
the `$\nu_2.\nu_3$' triangle which is the leptonic analogue of the
`$db$' triangle of the quark sector and is expressed as
 \be U_{e2} {U_{e3}}^* + U_{\mu 2} {U_{\mu 3}}^* + U_{\tau 2} {U_{\tau 3}}^*
=0\,.\label{n2n3} \ee Using the matrices constructed above and
following the same procedure as in the quark case, for the
`$\nu_2.\nu_3$' triangle we obtain the corresponding respective
values of $J_l$ as
 \be J_l= 0.009 \pm 0.003, \label{j-.05}\ee
 \be J_l= 0.018 \pm 0.006, \label{j-.1}\ee
 \be J_l= 0.024 \pm 0.009. \label{j-.15}\ee
Using the relation between $J_l$ and phase $\delta$, equation
(\ref{jd}), as well as considering the above values of $J_l$ and
 mixing angles to have Gaussian distributions, one can find the
corresponding histograms of $\delta$. Using the histograms, shown
in figure \ref{delneut}, the $\delta$ values corresponding to
$U_{e3}$ values $0.05 \pm 0.005$, $0.10 \pm 0.01$ and $0.15 \pm
0.015$ are respectively as follows
 \be \delta\simeq47^{\rm o} \pm 15^{\rm o}, \ee
 \be \delta\simeq45^{\rm o} \pm 15^{\rm o}, \ee
 \be \delta\simeq42^{\rm o} \pm 16^{\rm o}. \ee
It is interesting to note that the CP violating phase $\delta$
comes out to be around $45^{\rm o}$ and is not much sensitive to
$U_{e3}$ in the range $0.05-0.15$. The above calculated values of
$\delta$, indicating an almost $2.5\, \sigma$ deviation from
$0^{\rm o}$, are in line with the suggestion by several authors
\cite{marciano}, \cite{minakata1}, \cite{giunti} about the
expected CP violation in the leptonic sector. In particular, the
present analysis is broadly in agreement with a similar analysis
carried out by Farzan and Smirnov \cite{smirnov} and also with a
recent phenomenological analysis carried out by K.R.S. Balaji {\it
et al.} \cite{balaji}. Further, it is interesting to note that the
present analysis carried out purely on phenomenological inputs is
very much in agreement with several analyses based on expected
outputs from different experimental scenarios \cite{smirnov},
\cite{white}, \cite{marciano}, \cite{sugiyama}. In particular, the
analysis of Marciano and Parsa \cite{marciano} carried out for the
BNL-Homestake (2540 km) proposal is in complete agreement with the
present analysis in respect of expected error in $\delta$ and the
insensitivity of $\delta$ for values of $U_{e3}\gtrsim 0.05$.
Therefore, the BNL-Homestake experiment would not only shed light
on the existence of CP violation in the leptonic sector but would
also have implications for the scenario of Bjorken {\it et al.}
\cite{bjorken}.

To summarize, following analogy of the `$db$' triangle in the
quark sector, we have explored the possibility of the construction
of the `$\nu_2.\nu_3$' leptonic unitarity triangle using the MNS
matrix obtained by Bjorken {\it et al.} \cite{bjorken}. In
particular, modifying this matrix and considering values of
$U_{e3}$ suggested by different theoretical models we have
obtained the leptonic mixing matrices. Using these as well as the
`$\nu_2.\nu_3$' leptonic unitarity triangle, we have constructed
histograms for the Dirac-like CP violating phase $\delta$ in the
leptonic sector. Interestingly, from these histograms one can find
that for the $U_{e3}$ range $0.05-0.15$, there is an almost $2.5\,
\sigma$ likelihood of finding a non zero Dirac-like CP violating
phase $\delta$ with its central value to be around $45^{\rm o}$.
The present analysis is largely in agreement with the analysis of
Marciano and Parsa \cite{marciano} regarding the expected outcome
of the BNL-Homestake (2540 km) proposal.

  \vskip 0.5cm
{\bf Acknowledgements} \\ The authors would like to thank S.D.
Sharma, Sanjeev Kumar and Monika Randhawa for useful discussions
and the DAE, BRNS (grant No.2005/37/4/BRNS), India, for financial
support. G.A. would like to thank the Chairman, Department of
Physics, for providing facilities to work in the department.

\end{document}